\documentclass{egpubl}
\WsPaper
\electronicVersion

\ifpdf \usepackage[pdftex]{graphicx} \pdfcompresslevel=9
\else \usepackage[dvips]{graphicx} \fi

\PrintedOrElectronic

\usepackage{t1enc,dfadobe}

\usepackage{cite}

\usepackage{url}
\usepackage[usenames]{xcolor}
\usepackage{booktabs}
\usepackage{amsmath}
\usepackage{multirow}
\usepackage{tabularx}
\usepackage{array}

\title[A Space-Efficient Method for Navigable Ensemble Analysis and Visualization]
{A Space-Efficient Method for Navigable Ensemble Analysis and Visualization}

\author[A. Hota, M. Raji, T. Hobson, \& J. Huang]
{Alok Hota$^{1}$,
    Mohammad Raji$^{1}$,
    Tanner Hobson$^{1}$,
    and Jian Huang$^{1}$
    \\
    $^1$University of Tennessee, Knoxville, Tennessee, USA
}

\def\abstract{
    \@abstract}

\def\@abstract{\list{}{\leftmargin 0pc\rightmargin\leftmargin
        \parsep 0pt plus .1pt}\item[]{\textbf{Abstract}}\\\itshape}

\begin{document}

% make the title area
\maketitle

\begin{abstract}
Scientists increasingly rely on simulation runs of complex models in lieu of cost-prohibitive or infeasible experimentation. The data output of many controlled simulation runs, the ensemble, is used to verify correctness and quantify uncertainty. However, due to their size and complexity, ensembles are difficult to visually analyze because the working set often exceeds strict memory limitations. We present a navigable ensemble analysis tool, NEA, for interactive exploration of ensembles. NEA's pre-processing component takes advantage of the data similarity characteristics of ensembles to represent the data in a new, spatially-efficient data structure which does not require fully reconstructing the original data at visualization time. This data structure allows a fine degree of control in working set management, which enables interactive ensemble exploration while fitting within memory limitations. Scientists can also gain new insights from the data-similarity analysis in the pre-processing component.

\begin{classification} % according to http://www.acm.org/class/1998/
    \CCScat{Human-centered computing}{Visualization}{Visualization application domains}{Scientific visualization}
\end{classification}
\thispagestyle{empty}
\end{abstract}

\section{Introduction}\label{sec:introduction}

%
% stating the problem
%  - ensemble data is becoming common, but it is huge
%

%from Hank's talk
In today's science and engineering, simulations often replace cost-prohibitive or infeasible experimentation.
%reworded from previous draft
As complex computational models for such experimentations become prevalent, the resulting ensemble data output are more easily available than before.
These datasets are important for verifying stability of predictions, quantifying uncertainty, and evaluating the sensitivity of key parameters, and provide a key role in the reproducibility of science.
Due to the large number of simulation runs that are typically executed, the resulting ensemble dataset has an extra layer of complexity beyond the usual volumetric and time-varying aspects of the data.
As a result, these datasets are very large and challenge current visualization algorithms and systems.

%
% characterize the problem
%

Ensemble visualization and analysis is a long-standing problem, and a wide variety of solutions have been proposed.
For example, frameworks have been developed for specific domain science applications~\cite{Sanyal2010}\cite{Hollt2014}\cite{Biswas2016}, as well as ensemble analysis targeting general cases~\cite{Potter2009}\cite{Obermaier2016}.
%requirements, haven't seen the data
%While these frameworks are highly informative and interactive, they perform analysis-based ensemble visualization.
%This requires that users must already have knowledge about the dataset or have preconceived hypotheses, which stem from seeing the data in the physical space, before being able to interact with the data.
Visual exploration is a critical phase in the data analysis process that allows users to make the hypotheses needed for analysis.
However due to I/O and memory bottlenecks on personal machines, interactive exploration has been challenging.
%To the author's knowledge, existing tools do not provide exploration-based ensemble visualization that allow interactive volume rendering of ensemble members.

We created three defining guidelines for an exploratory ensemble visualization application based upon a natural use-case.
Such a tool must: (i) be able to visualize a single run over all time steps, (ii) allow switching between runs at will, and (iii) be able to visualize agreement across multiple runs.
%These three requirements characterize exploratory ensemble visualization, and highlight where current methods are insufficient.
Current software packages, such as VisIt and ParaView, present a time-tested and intuitive interface for exploration and fulfill the first requirement.
However they may incur costly loading times when switching between runs (the second requirement).
%These requirements can be more generally defined as the ability to visualize an ensemble dataset with minimal memory footprint and without interruption to the interactive exploration process.

In this paper, we present NEA, an exploratory ensemble analysis and visualization tool that significantly reduces the spatial complexity of ensemble datasets to enable interactive visual exploration and analysis.
NEA takes advantage of the expected data similarity between runs of an ensemble and between time steps within a run.
The data is stored in a new spatially-efficient data structure that does not require full dataset reconstruction at visualization time.
The data structure allows for a fine degree of control over the visualization working set, allowing the data to more easily fit within memory limitations.
The data structure also fits well with disk cache, accelerating load times while traversing an ensemble.
This in turn enables interactive visualization using existing rendering methods.
%This spatial efficiency allows for data to more easily fit within memory limitations, which enables interactive analysis and visualization using existing methods.
Additionally, the data similarity analysis done by NEA's processing step allows for a fast ensemble agreement analysis.

We describe the methodology for computing the data structure in Section \ref{sec:method}.
In Section \ref{sec:results}, we demonstrate the efficacy of NEA's reduction using a Weather Research and Forecasting (WRF) superstorm ensemble using various parameters.
The ensemble itself is described in Section \ref{sec:res-ensembles}.
We then show that NEA produces faithful recreations of the original data at visualization time on a class of machine smaller than the status quo for large-scale visualization.
We demonstrate how input parameters during the processing component affect the results, and we provide a profiling tool based on heuristics to aid in parameter selection.
We conclude our findings in Section \ref{sec:conclusion}.
%We showcase the ability to manage the working set with a simple VTK-based visualization tool that allows for interactive exploration of the ensemble on a class of machine smaller than the status quo for large-scale visualization.

\section{Related Works}

Balsa Rodriguez et. al~\cite{BalsaRodriguez2014} have begun characterizing the task of volumetric data reduction and have begun a taxonomy of the process.
The various compression techniques surveyed show that working set management, or an ``adaptive working set,'' allow for large-scale visualization in restricted memory conditions in the context of GPUs.
Additionally, data segmentation, transformation, and restructuring have been shown to aid in visualizing time-varying volume data.
However we present our work with some key differences and considerations.

The first is that we put emphasis on and target ensemble data.
These datasets present an issue of scale in that there are now multiple sequences of large, time-varying volumes.
Visual exploration of these datasets is difficult due to their size.
Second, we target general-use computing hardware used for visualization, such as personal laptops.
%Small scale machines, such as personal laptops, commonly do not have discrete GPUs and instead rely on onboard graphics or CPU-based solutions.
Even on these machines, today's ``Little Iron,'' high-quality direct volume rendering is tractable with CPU solutions.

\subsection{Direct Volume Rendering}

Proven methods exist for volume rendering, such as shader implementations on GPUs and fast CPU-based ray tracing libraries like Intel's OSPRay \cite{ospray}.
Software packages like VisIt \cite{visit} and ParaView \cite{paraview} present an interface for these powerful rendering solutions, allowing for interactive volume rendering.
However, regardless of the number of GPUs or multi-/many-core processors present, the main limitation for ensemble datasets is memory.
On large-scale machines, memory may not be a bottleneck, but such machines have restricted access and an associated utilization cost.
It is difficult to manage and maintain the ensemble analysis and visualization working set to fit within the much more stringent memory limitations of commodity machines.
By controlling the working set, simple existing rendering methods--such as VTK's built-in CPU ray casting renderer~\cite{vtk}--can be used for interactive ensemble exploration.

\begin{figure*}
    \centering
    \includegraphics[width=\linewidth]{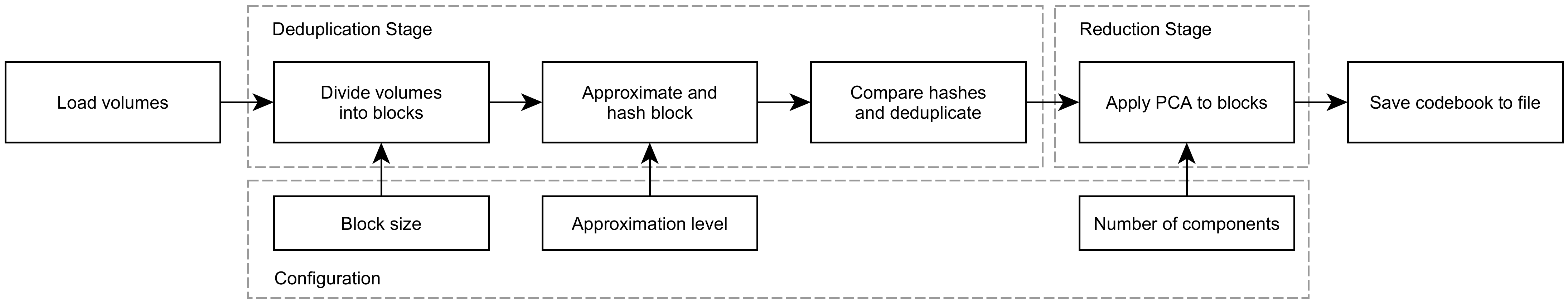}
    \caption{An overview of the NEA processing pipeline. The two processing stages are computed in sequence with three user input parameters controlling different aspects of the pipeline. The final output is a codebook file that can be read by the visualization component.}
    \label{fig:processing-pipeline}
\end{figure*}

\subsection{Ensemble Analysis and Visualization}\label{sec:bg-ensemble}

The use case for ensemble analysis and visualization is very broad, as it applies to any domain science work that studies or uses simulation data.
Researchers have applied ensemble analysis to specific domain science applications, such as ocean simulation analysis~\cite{Hollt2014}, weather and climate prediction and analysis~\cite{Sanyal2010}\cite{Biswas2016}, and particle physics analysis~\cite{Lin2012}.
On the other hand, more generalized ensemble visualization frameworks have been developed.
These applications and frameworks enable efficient and detailed agreement or trend analysis~\cite{Obermaier2016}\cite{Demir2014}\cite{Chen2015}\cite{Potter2009}.
Different methods of efficiently comparing ensemble members have been developed as well, such as using hierarchical clustering~\cite{Hao2015}.
These applications also display a variety of visualization methods, include isosurfaces colored by a scalar, spaghetti plots and glyphs, heat maps, two-dimensional projections~\cite{Ahrens2010}\cite{Bensema2015}, or newly-created visualization methods.
Ensemble analysis and visualization has been succinctly characterized by Obermaier and Joy \cite{obermaier2014}.

These applications are highly informative and provide complex analysis useful for scientists.
However exploration of ensemble data has proved a challenge, due in large part to memory and I/O limitations on the machines used for visualizations.
These bottlenecks prevent large volumes being held in memory, or make loading and transferring the data impractical.

\subsection{Large-Scale Visualization}\label{sec:bg-scale}

The current status on visualization and analysis machines, dubbed the ``Little Iron,'' suggest that they exist as a percentage of the large-scale systems, the ``Big Iron''\cite{Bethel2011}.
The status quo for visualization systems is approximately 10\% of the Big Iron, the machine on which simulations are run.
Big Iron systems have restricted access and are in general expensive to utilize.
Utilization costs can be described in terms of currency or CPU hours, both of which can be quickly exhausted by interactive remote visualization sessions.

\textit{In situ} visualization approaches this by not saving data in lieu of visualizing while the data is in memory during simulation time~\cite{Bauer2016}.
VisIt and ParaView provide Libsim~\cite{Whitlock2011} and Catalyst/Cinema~\cite{Ahrens2014}, respectively, to allow using the applications for \textit{in situ} visualization.
By providing a set of images, users could interactively explore ensembles without the necessary requirements for storage or memory.
However, post-hoc visualization and analysis is still critical to many scientists.
For example, proposals to the National Science Foundation (NSF) must include a data management policy, outlining how data will be stored and accessible~\cite{nsf-data}.
That is, any data resulting from a simulation must be stored and available for use after the original work has been completed.
The peer-accessibility of such data is critical to the reproducibility of scientific works.

We define a ratio between the performance of the simulation machine to the visualization machine as the \textit{machine ratio}, or MR.
The MR could represent the relative performance in terms of CPU cores, GPU(s), memory, or a combination of the three.
%A key aspect to the NEA pipeline is that the resultant codebook can be used to visualize an ensemble on commodity hardware, such as a laptop.
On personal machines--such as laptops--the most limiting resource is memory, so a memory-based MR is most appropriate.

The superstorm simulation was run on a cluster with 4 TB of memory distributed among 512 nodes.
If visualizing on the same machine, the MR would be 1:1.
If the MR were 10:1, i.e. the status quo, the visualization machine would contain roughly 400 GB of memory.
Standard laptops today come with memory ranging between 4 and 16 GB.
Section \ref{sec:res-perf} demonstrates performance on a laptop with 8 GB memory, or an MR of 512:1.
By shrinking the requirements of the visualization machine, large ensemble data can be more accessible for reproducibility and further study.

\subsection{Hierarchical Structuring}

Methods for restructuring data into a hierarchical representation have been used for large volumes.
These techniques build hierarchical structures utilizing temporal coherence~\cite{Jang2012}\cite{vmv.20151255}, or spatial and temporal features~\cite{Linsen2002}\cite{Visualization2005}\cite{Shen1999}.
Additionally, structures may be built around features in histograms~\cite{Ip2012}.
Temporal coherence may not necessarily scale well across the different runs of an ensemble dataset.

Hierarchical structures enable providing details on demand during rendering.
Similar to level-of-detail models for polygonal meshes, the amount of detail provided increases as the camera nears the object.
Level of detail may also vary as a function of position within the volume itself~\cite{Guthe2002}.
A dictionary-based hierarchical method optimized for GPU decompression and rendering has been studied as well~\cite{Gobbetti2012}.

These out-of-core techniques provide a structure representing the data at various levels.
%The overhead of different layers is mitigated in most cases with a reduction technique, such as wavelet transformation.
However, our approach does not require building these trees at varying levels.
By applying deduplication, we can remove repeated data without the need to remove fine details at any resolution.
Additionally, once a volume is reconstructed during visualization time, users are free to interact with the volume with no additional loading or processing required.

\subsection{Data Reduction and Compression}

A common tactic for lossless data compression is deduplication, which segments the data and replaces duplicate segments with a reference to an original copy.
This is a known technique used often in the data storage and encryption communities~\cite{Paulo2014} for everyday workstation storage efficiency~\cite{meyer2012study} and cloud storage~\cite{stanek2014secure}\cite{Srinivasan2012}.
This technique can be applied to scientific ensemble data by dividing volumes into blocks and removing repeated blocks across the ensemble.
This leverages the expected data similarity between runs.
Additionally, this provides more control over managing the working set, as individual blocks can be loaded as needed instead of whole volumes.

Data segmentation for working set management has been applied to data for visualization purposes in previous works, where segments are uniform~\cite{Fout2005}\cite{Ning1993} or based on features~\cite{Bremer2011}.
However these approaches do not target ensemble volumes, nor do they apply deduplication to the divided segments.

Lossy data compression can be used with scientific data as well.
Compression at the floating-point level is possible~\cite{Lindstrom2014}.
Vector quantization~\cite{Fout2005}\cite{Ning1993}\cite{Schneider2003}
and wavelet-based transformations~\cite{Ihm1998}\cite{Rodler1999} have been used to reduce the footprint of volume data as well.
Guthe and Stra\ss{}er \cite{Guthe2002} in particular show that wavelet transformation for volume data followed by entropy encoding and run-length Huffman encoding provides fast compression and decompression with high compression ratios.
Combined with a fast method for calculating a compression threshold~\cite{donoho1994ideal} shows that data can be quickly compressed and later decompressed at visualization time.

These methods of data reduction can be employed at different levels of data hierarchy that exist in an ensemble dataset.
We show that deduplication can be used to exploit large scale similarity features across the entire ensemble to reduce the overall data representation into a set of unique blocks.
Conventional data compression techniques, such as vector quantization through principal component analysis or wavelet transformation, can then be used to reduce the smaller-scale features present at the block level.

\section{Methodology}\label{sec:method}

NEA consists of two main components: a one-time processing component that generates a \textit{codebook} for a given ensemble, and a visualization component that intelligently manages the working set while rendering.
The processing component is performed in two stages, described in Sections \ref{sec:method-dedup} and \ref{sec:method-pca}.
The first is a serial Python application that analyzes the ensemble and performs deduplication.
The second stage performs a dimensionality reduction to the deduplicated data and saves the reduced data as a codebook.
%We compared the use of principal component analysis (PCA) and discrete wavelet transformation (DWT) for reduction.
Our implementation leverages principal component analysis (PCA) for reduction, but this choice is modular and other methods can be used instead.

The overall processing pipeline is described in Figure \ref{fig:processing-pipeline}.
The new codebook data structure contains grids that define the volumes of an ensemble, and a block lookup table (a hash table), that contains a mapping between block IDs and data.
The visualization component, described in Section \ref{sec:method-vis}, streams data from the codebook when volumes are requested by the user.

\subsection{Ensemble Characteristics}\label{sec:method-ensemble}

\begin{figure}
\centering
\includegraphics[width=0.7\linewidth]{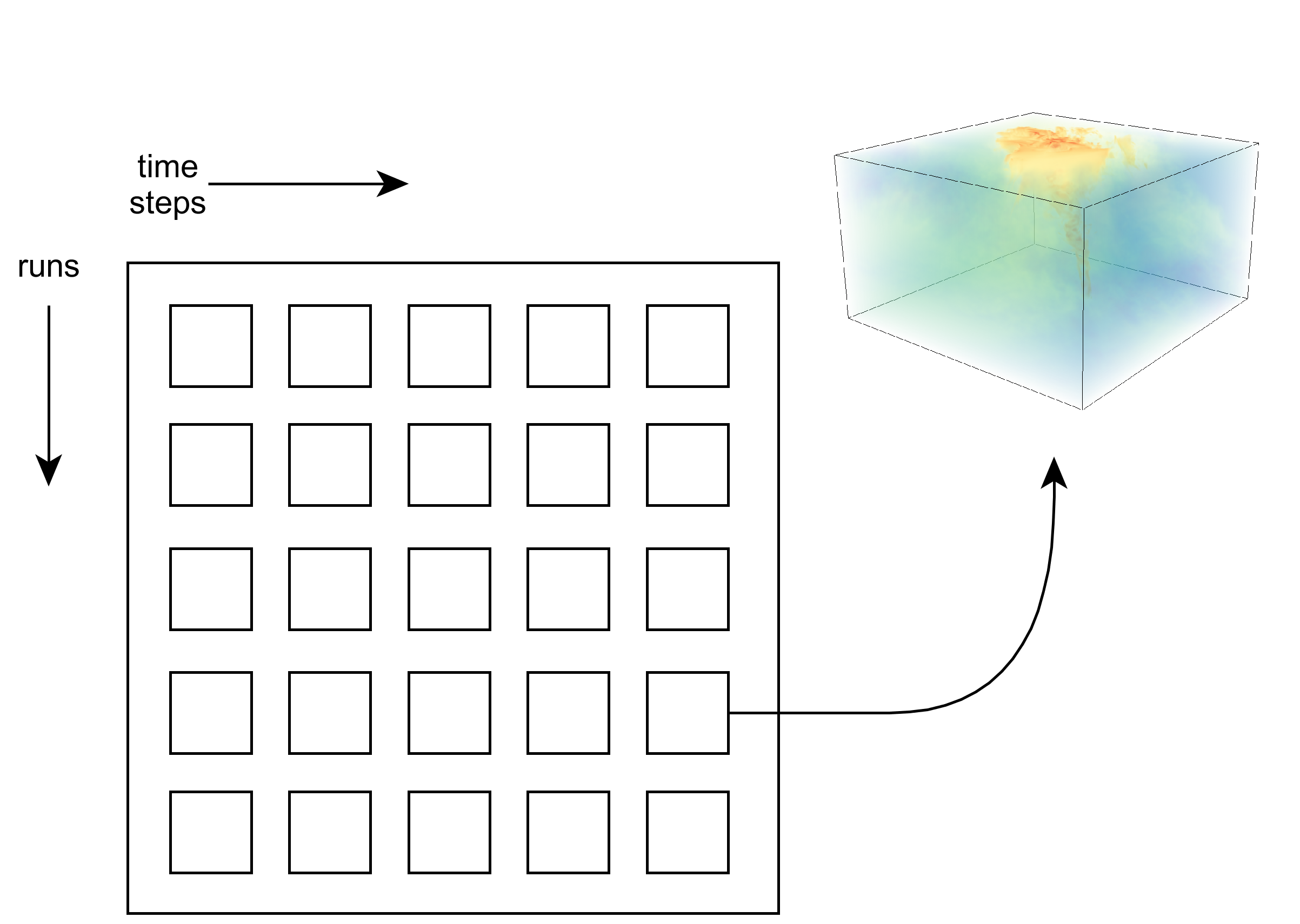}
\caption{Conceptual depiction of an ensemble. The vertical axis represents different runs of the simulation, while the horizontal axis represents the timesteps. Thus each element $E_{rt}$ is a single volume from the ensemble. Similarity is expected across runs, as they are the result of the same model. Similarity between subsequent timesteps in a run is governed by the temporal granularity of the data and by the variables' rate of change.}
\label{fig:ensemble-matrix}
\end{figure}

Ensemble datasets can conceptually be described as a two dimensional grid such that each row is an ensemble member, or \textit{run}, consisting of several timesteps.
This \textit{ensemble space} is illustrated in Figure \ref{fig:ensemble-matrix}.
Supposing an ensemble has $R$ runs, each containing $T$ timesteps, then each element $E_{rt}$, for $r\in[1, R]$, $t\in[1, T]$,
of this grid is a single volume from the ensemble.
This conceptual model is useful for understanding the characteristics of an ensemble, as well as understanding the scope of ensemble analysis and visualization.

A typical process applied to ensemble datasets is agreement; that is, whether multiple simulation runs agree with one another.
In the context of the two dimensional ensemble grid, this can be described as comparing rows against one another.
That is, the elements $E_{rt}$ for $r\in[1, R]$ are expected to be similar.

\begin{figure}
    \centering
    \includegraphics[width=0.4\textwidth,trim={0 0 0 0},clip]{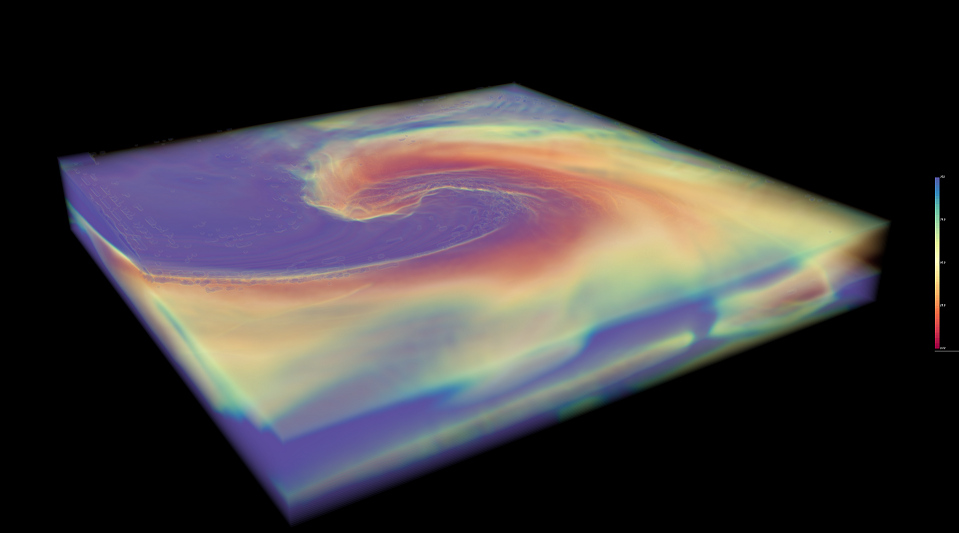}
    \caption{An example of the NEA visualization component. This shows run 1, timestep 30 of the ensemble. Blocks are loaded as needed, keeping the memory usage low, when visualizing the ensemble.}
    \label{fig:vis-example}
\end{figure}

\subsection{Deduplication}\label{sec:method-dedup}

The general process for deduplication is (i) divide data into blocks, (ii) create a unique identifier for each block, and (iii) remove duplicated blocks and replace them with references to identifiers.
These three steps are shown in Figure \ref{fig:processing-pipeline} in the Deduplication Stage.

To make comparisons efficient, volumes are divided into uniformly-shaped blocks of a user-defined size $x\times y\times z$, the \textit{block size}.
The resulting \textit{grid} has dimensions equal to the ceiling of the original volume's dimensions divided by the block size.
If the original volume's dimensions are not a multiple of the block size, blocks along the edge will contain fill values as padding.
For example, the superstorm dataset contains volumes of size $254\times 254\times 37$.
With a block size of $4\times 4\times 4$, the resulting grid would be of size $64\times 64\times 10$.
Because none of the dimensions are a multiple of 4, blocks along the far edges of each dimension will be padded.
Each block is given a five-dimensional coordinate $(r,t,i,j,k)$ where $r, t$ is a volume's location in the ensemble space, and $i, j, k$ are the block's coordinates within that volume's grid.
This coordinate defines a unique location in the ensemble, and serves as a block ID.

Comparing blocks against one another by iteratively comparing every floating-point value within them would be a time-consuming $O(n^3)$ algorithm.
Instead, we flatten the block's data into a one-dimensional vector, generate a hash value for the block using the SHA256 algorithm, and compare the resulting hash values to one another.
To mitigate the hash algorithm's sensitivity to input values, we first round the data in the flattened vector.
The decimal place to which values are rounded can be chosen by the user.
Looser decimal place restrictions increases the likelihood of multiple blocks reducing to the same hash (i.e. the blocks contain equivalent data), but also decreases the accuracy of the final data representation.
Hashed blocks are then compared one-to-one across the entire ensemble.
This allows for checking for similarity between runs, where it is expected, as well as between timesteps.
Additionally, this can catch similarity between blocks that are both spatially and temporally distant.
After comparing hashes, we obtain a mapping between a single hash value to one or more block locations.

A single block is chosen from each group of matching blocks to represent the group in the deduplicated dataset.
This block is given a canonical ID and its data is stored in the block lookup table.
%The data corresponding to the chosen block is loaded, and the block is given a new canonical ID.
All other blocks in the group receive the same canonical ID, mapping similar blocks across the ensemble to a single copy of the data.
Finally, the codebook data structure is generated containing a block lookup table and a set of grids.
The block lookup table maps canonical IDs to data from the dataset, and describes the \textit{minimum representative set} for the ensemble--the smallest set of unique blocks that can be used to recreate any volume in the ensemble.
The grids in the codebook describe which block should be loaded into which location to rebuild any given volume in the ensemble.

\subsection{Data Reduction}\label{sec:method-pca}

%% placed this here so that it shows up on the next page
\begin{figure*}
    \centering
    \includegraphics[width=0.49\textwidth]{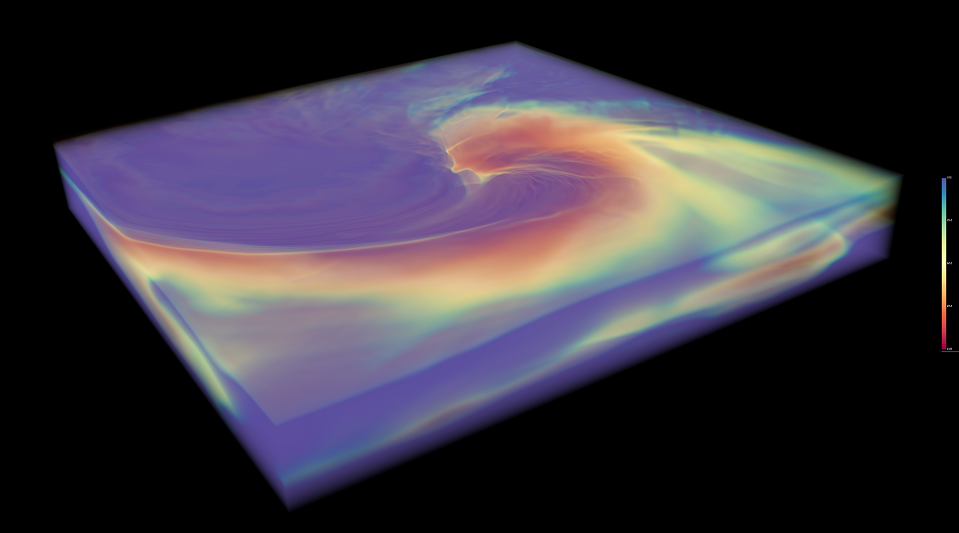}
    \includegraphics[width=0.49\textwidth]{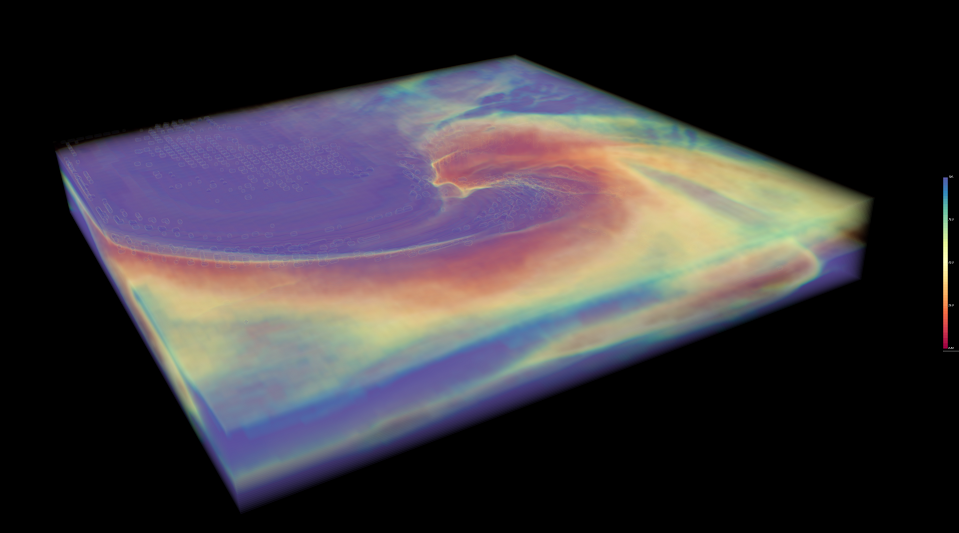}
    \caption{Comparison screenshots from the NEA visualization component. The original volume from run 1, timestep 39 is shown on the left, while the right shows the same volume loaded from a codebook. The parameters chosen for the codebook were designed to cause many blocks to be deduplicated and were then heavily compressed. The result was a codebook almost 7x smaller than the original that has noticeable artifacts, but still faithfully shows all large-scale and some small-scale features.}
    \label{fig:comparison}
\end{figure*}

To further minimize the footprint of the codebook on disk, we apply a known data reduction technique.
We tested an implementation of principal component analysis (PCA) to compress data at the block level.
However other methods can be used as well, as shown in Section \ref{sec:res-wavelet}.
%To further minimize the footprint of the codebook, we apply PCA as a data reduction technique.
%The choice of PCA is not critical, and its usage in the pipeline as a reduction or compression element can be replicated with other algorithms, such as wavelet transformation.
A trade-off to this step is that choosing fewer components for greatly reduced data representation results in lower accuracy.
Since both methods present lossy compression, values after performing the inverse transformation may have missing information or stray outside the original bounds of the variable.
%Values after performing the inverse PCA transformation may stray outside of the original bounds of the variable.

\subsubsection{PCA}
The blocks of data stored in the block lookup table are used as input for PCA.
Each block is transformed into a one-dimensional vector.
In vector representation, each block can be considered as a point in an $n$-dimensional coordinate system, where $n$ is the product of the block size dimensions.
Users can choose any number of components in $[0, n]$ for transformation.
After PCA is applied, the block lookup table in the codebook holds a map from unique block IDs to a vector of transformed data.
%The length of the vectors is the number of components chosen for PCA.
Finally, the basis vectors are saved alongside the transformed codebook as a metadata file.

\subsection{Visualization}\label{sec:method-vis}
The NEA visualization component presents a simple graphical interface for exploring a given codebook.
The interface is built with VTK's Python bindings. % to ease development.
It uses VTK's built-in CPU-based ray casting renderer for volume rendering.
%Several screenshots from the tool are shown in Figure \ref{fig:teaser} comparing codebooks with different user input parameters.

When initialized, the tool presents users with the volume from the first run and first timestep of the dataset.
Users are able to move forward and backward through the timesteps of a run, as well as switch between runs, with keyboard interaction.
%Users are able to move forward and backward through the timesteps of a run using the left and right arrow keys on the keyboard.
%Similarly, users can switch between runs using the up and down arrow keys.
This allows the NEA visualization tool to remain consistent with the conceptual ensemble space shown in Figure \ref{fig:ensemble-matrix}.
%A selection of color maps are provided, including Cool to Warm, Spectral, reversed Spectral, and Viridis.
%The user can cycle between these by pressing the C key.
An example of the simple interface presented is shown in Figure \ref{fig:vis-example}, which shows a codebook with the following parameters: $16\times 16\times 16$ block size, 0 decimal place approximation (i.e. integer rounding), and 512 components for PCA (1/8 of maximum possible).

During visualization, the tool can leverage NEA's codebook structure to manage the working set with a fine degree of control.
The codebook allows for single blocks to be streamed in without loading the entire set of blocks into memory.
As blocks are loaded, they can be inversely transformed using the PCA basis vectors.
%Depending on the method used, the blocks can be inverse transformed as they are loaded, either using the PCA basis vectors, or by run-length decoding followed by inverse wavelet transform.
The visualization tool thus maintains two main data structures: (i) the current volume's grid describing which block ID contains the data for a certain region, and (ii) a table mapping only the current grid's block IDs to data (a subset of the full block lookup table).

At any one point, the visualization tool is holding the data for a single grid by using a set of current block IDs, $CB$, held in a table.
When switching volumes, NEA loads the grid for the new volume, which contains a set of new block IDs, $NB$.
The set of blocks to keep in memory ($k$), the set of blocks to be loaded in ($l$), and the set of blocks to be safely discarded from memory ($d$), are defined as follows.
%is determined by which block IDs are in common between $CB$ and $NB$, i.e. $k = CB\cap NB$.
%The set of blocks that need to be loaded in, $l$, can be described as $l = NB - k$, and the set of blocks to be safely discarded from memory, $d$, is therefore $d = CB - k$.
\begin{align}
k &= CB\cap NB\nonumber\\
l &= NB - k\\
d &= CB - k\nonumber
\end{align}
This computation is done whether switching timesteps or runs, as both movements through the ensemble are logically equivalent.
PCA inverse transformation is applied to data loaded from blocks $l$.
Once the data has finished loading, the new volume is rendered.
The user can interact with the volume with the typical rotation, zooming, and panning.
Block loading only occurs once when a volume is loaded, and not during interaction.

Another benefit of the codebook's block streaming capability is it simplifies the process of ensemble agreement visualization.
At any timestep, the user can press the A key to see agreement for that timestep across all runs.
That is, viewing agreement for a column in the conceptual ensemble grid.
For agreement visualization, block data need not be loaded into memory.
Because blocks have already been compared, the visualization tool only needs to detect which IDs occur in each location of a grid.
Grids are much smaller than the original volume, and thus require much less time for comparison.

\begin{figure*}
    \centering
    \includegraphics[width=0.49\linewidth]{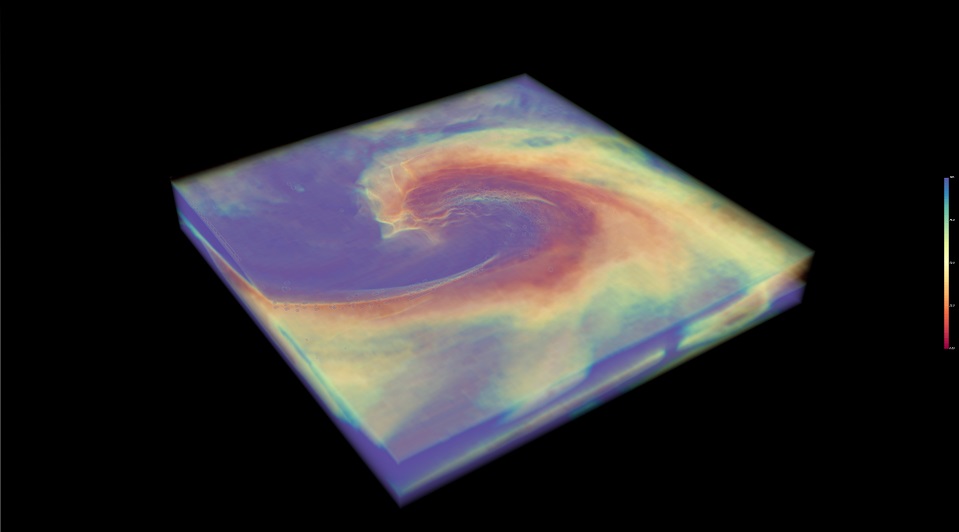}
    \includegraphics[width=0.49\linewidth]{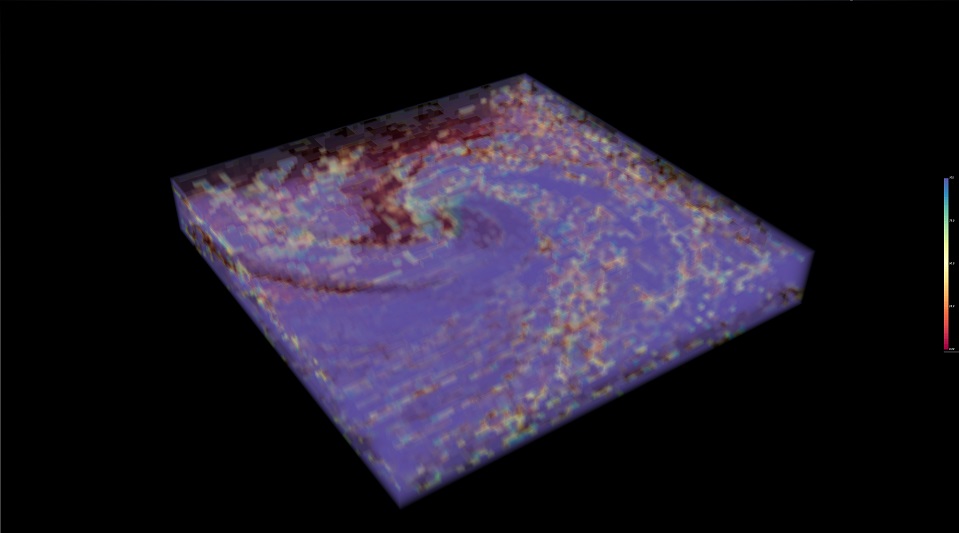}
    \caption{Visualizing agreement across runs. The left image shows the volume at run 3, timestep 45. This is one of the more turbulent timesteps for the superstorm simulation. Still, the ensemble exhibits a high degree of agreement, shown on the right. Areas around edges of the storm front disagree while the large area within the storm contains similar values across runs.}
    \label{fig:agreement}
\end{figure*}

This process is straightforward.
The current volume, located at some run $r_c$ and timestep $t$, is used as a reference point.
The grids for all volumes in runs $r\in R$ and time step $t$ are loaded, and a ``running total'' grid is initialized.
The block ID at each location $i,j,k$ is compared between each grid and the reference point.
For every match, the value in location $i, j, k$ in the running total grid is incremented.
After each grid has been compared, the sums in the running total grid are divided by the total number of runs $R$ to show agreement as a percentage of runs with the same block IDs.
Figure \ref{fig:agreement} shows a screenshot of the visualization component displaying agreement with run 1, timestep 48 as the reference.

\section{Performance and Results}\label{sec:results}

We tested the NEA processing and visualization components.
We chose parameters to first show how the processing component reduces data size for a large ensemble in Section \ref{sec:res-ensembles}, and then tested the resultant size's sensitivity to input parameters in Section \ref{sec:res-sensitivity}.
Section \ref{sec:res-perf} shows the visualization component's performance and effectiveness at managing the working set and the trade-offs between input parameters, as well as a comparison with existing tools.
Due to these complex trade-offs, we created a profiling script that estimates ``good'' values for the input parameters by sampling the data, described in Section \ref{sec:res-profile}.
Using these values, users do not need to conduct a trial-and-error parameter space search.

\begin{table}
    \centering
    \begin{tabular}{c c c}
        \toprule[1pt]
        Block & Decimal & PCA \\
        Size & Places & Components \\
        \midrule
        $4\times 4\times 4$ & -1 & $1/2$ \\
        $8\times 8\times 8$ & 0 & $1/4$ \\
        $16\times 16\times 16$ & 1 & $1/8$ \\
        $8\times 8\times 1$ & 2 & $1/16$ \\
        \bottomrule[1pt]
    \end{tabular}\\
    \textit{List of all possible PCA components:\\
    4, 8, 16, 32, 64, 128, 256, 512, 1024, 2048}\\
    \caption{The parameters modified for testing the NEA processing component. These choices cover a wide spectrum of the possible configurations for NEA. The choices for number of PCA components are listed as fractions because the number of possible PCA components is dependent on the number of elements in a chosen block size. For example, given a block size of $8\times 8\times 8$, there are 512 elements. The number of PCA components then varies between $512/2 = 256$, $512/4 = 128$, and so on. Computing this for each block size results in the list given above. Each combination of the parameter choices was tested, resulting in 64 test codebooks for each data reduction type.}
    \label{tab:parameters}
\end{table}

\subsection{Large Ensembles}\label{sec:res-ensembles}

The superstorm ensemble depicts a storm covering the eastern half of the United States between the 12th and 13th of March, 1993~\cite{Sanyal2010}.
The ensemble consists of 112 runs containing 49 timesteps each.
Each volume contains 114 variables of differing dimensionality.
We extracted the three-dimensional humidity variable, which has an extent of $254\times 254\times 37$.
We processed a segment of the extracted dataset consisting of all timesteps in 40 runs, or 1960 total volumes.
The original size for this segment of the dataset was 17.4 GB; each extracted volume was 9.1 MB.

To test processing, we ran with following configuration:
a $4\times 4\times 4$ block size, -1 decimal place approximation (i.e. rounding to the nearest tens), and 8 PCA components.
The small block size combined with high tolerance for matching was chosen to push the algorithm to greatly deduplicate and compress the data.
The deduplication processing stage took 7 hours, 22 minutes, and 41 seconds on a machine with 2x Intel Xeon E5-2670 CPUs at 2.6 GHz and 256 GB memory.
The following PCA application stage took 7 minutes and 8 seconds.
During processing, maximum memory usage occurred during the PCA stage, using 77 GB.
After processing, the resulting PCA codebook was 2.5 GB in size, a 6.96x reduction in size.
A comparison between the original data and the codebook data for the volume at run 1, timestep 39 is shown in Figure \ref{fig:comparison}.
Note that although compression artifacts exist for the regenerated volume shown, all large-scale (and some small-scale) features across the ensemble are still presented fairly accurately despite being stored at almost 14\% the original size.

\subsection{Parameter Sensitivity}\label{sec:res-sensitivity}

To test the processing pipeline's sensitivity to input parameters, we created a set of options that target a range of possible configurations.
The user has direct control over three parameters: (i) the block size used when dividing volumes and deduplicating, (ii) the number of decimal places rounded before hashing, and (iii) the number of components used in PCA.
Table \ref{tab:parameters} shows the four possible values for each parameter, resulting in 64 total test configurations for each method of data reduction.
The number of components for PCA is a function of the chosen block size.
That is, for a given block size $a\times b\times c$ where $n=abc$, $n$ is the maximum number of available components in PCA.
Any value chosen for the number of components to keep after PCA must be $\leq n$.
For testing, we chose four fractional values of $n$ that represented realistic choices of the number of components to keep.
For example, with a block size of $4\times 4\times 4$, $n=64$ and a PCA component choice of $1/2$ results in $64/2=32$ components kept.

\begin{figure}
\centering
\includegraphics[width=0.9\linewidth]{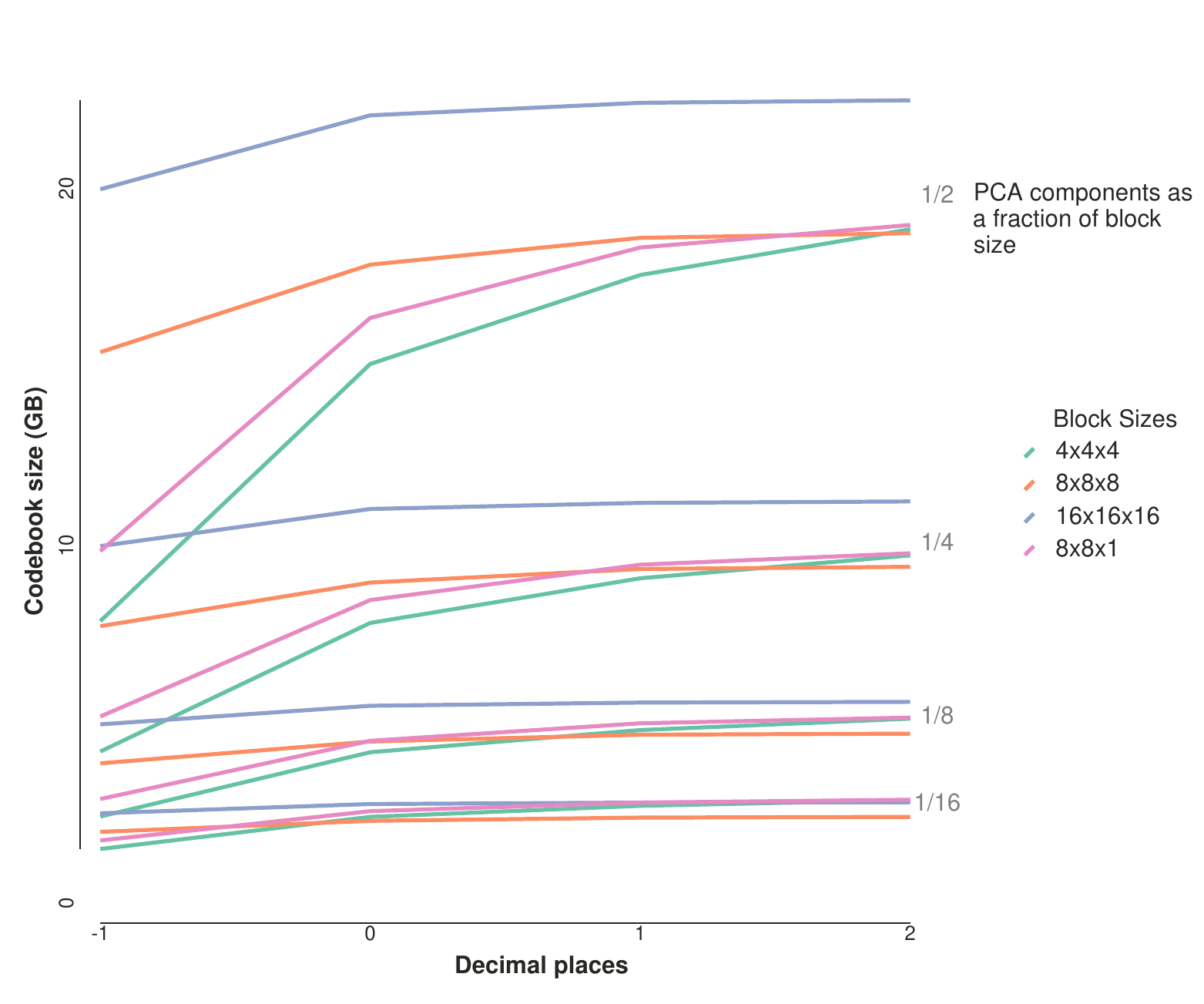}
\caption{Plot showing resultant codebook size as a function of block size, decimal place, and number of PCA components. Block size is shown by color. For each block size, four lines are drawn showing the number of PCA components kept. As the number of components kept is reduced, the overall effect of both block size and decimal place are less important.}
\label{fig:codebook_size}
\end{figure}

There is a set of important trade-offs between the different user parameters and the resultant codebook size and visualization performance.
Figure \ref{fig:codebook_size} summarizes these trade-offs.
Smaller block sizes deduplicate more effectively.
This is straightforward as smaller block sizes require fewer values to be compared.
The number of decimal places used for approximating before hashing affects the degree of deduplication as well.

However, the relative effect of both block size and decimal place on the final codebook size is less pronounced as the number of PCA components is reduced.
The final codebook size is more closely linked to the number of components than the other two parameters.
This is shown by the naturally grouped lines in Figure \ref{fig:codebook_size}.

While smaller block sizes deduplicated better and ultimately provided the smallest codebooks, they require a much denser grid representation for rebuilding volumes.
For example, with a block size of $4\times 4\times 4$, there are $64\times 64\times 10=40960$ blocks per $254\times 254\times 37$ superstorm volume (see Section \ref{sec:method-dedup}).
Each block requires a 4-byte integer for holding a block ID, totaling 160 KB per volume.
With a set of 1960 volumes as described in Section \ref{sec:res-ensembles}, this results in 306.25 MB for holding the grid alone.

The effect of this is that initial reads from the memory mapped codebook are slow for small block sizes, as it seeks through the table to access many small bits of data.
Data loading is gradually accelerated by disk cache.
On the other hand, larger block sizes have must faster access times at start, but are not accelerated as much by disk cache.

\subsection{Visualization Performance}\label{sec:res-perf}

%\begin{figure*}
\begin{figure}
    \centering
    %    \hspace*{2em}
    \includegraphics[width=0.45\linewidth]{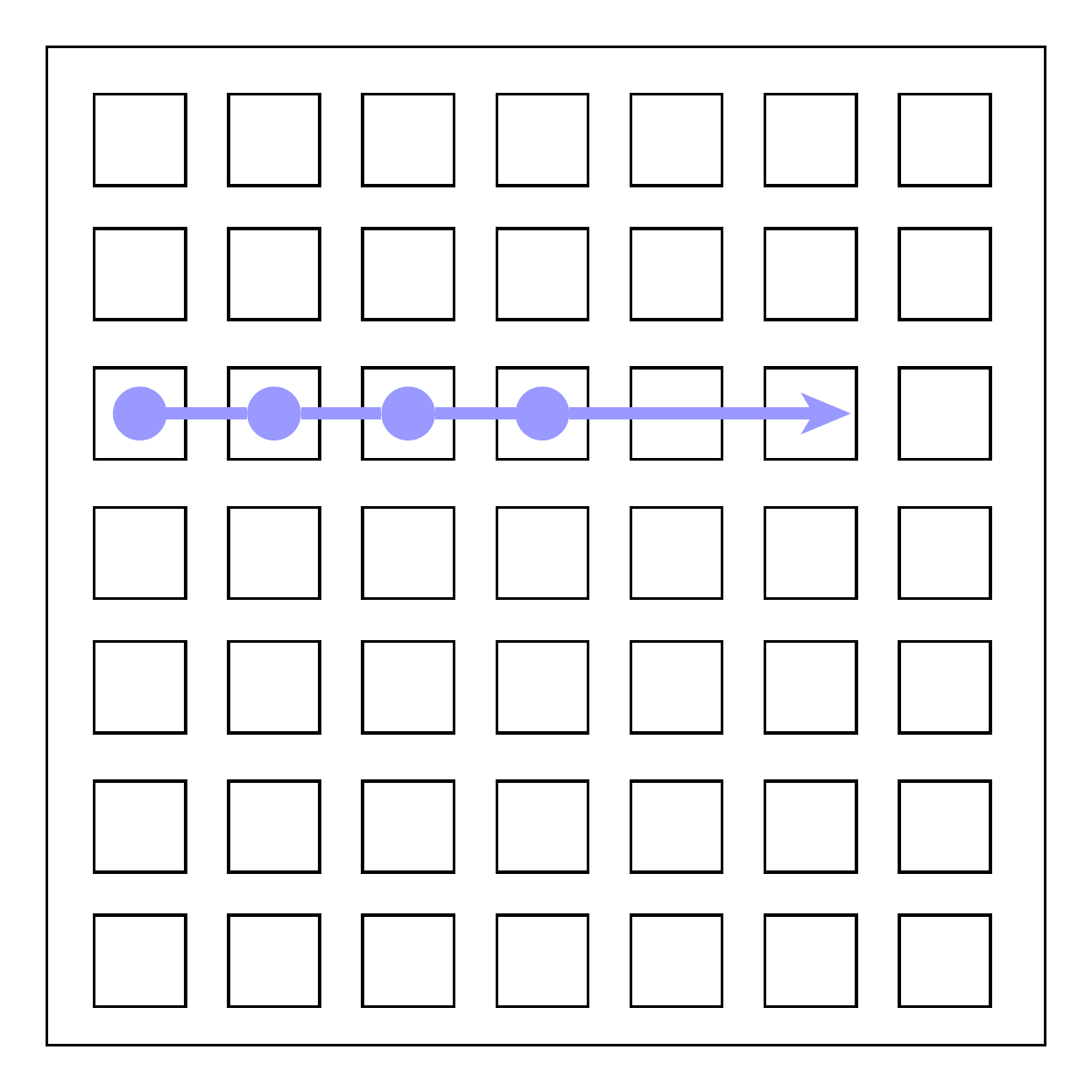}
    %    \hspace{2em}
    \includegraphics[width=0.45\linewidth]{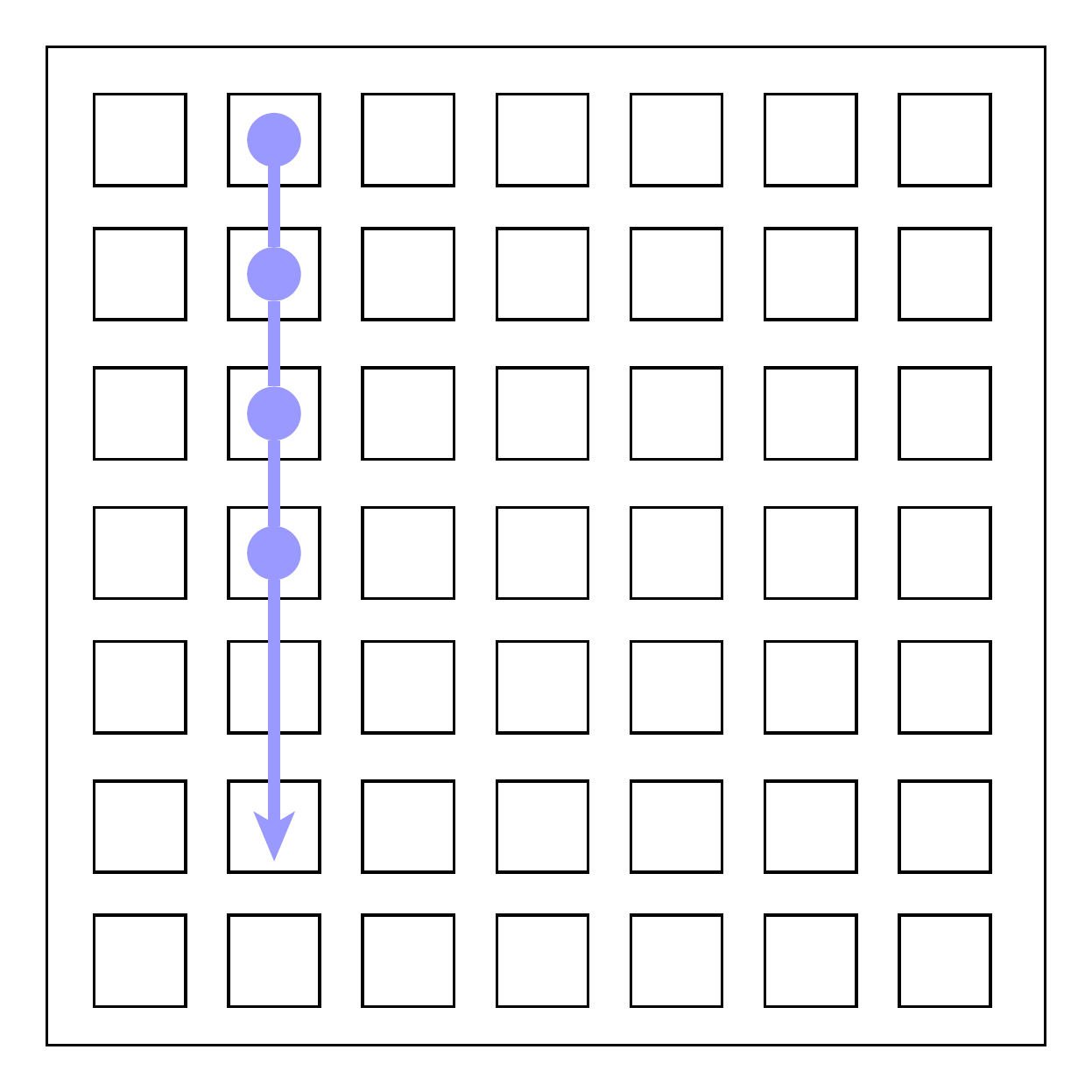}
    \caption{The visualization performance tests used these patterns to simulate different methods of exploring an ensemble. The left pattern is visualizing sequential time steps in a single run, while the right is visualizing one time step across multiple runs.}
    \label{fig:vis-test-patterns}
\end{figure}
%\end{figure*}

We measured the performance of the visualization component when rendering four codebooks selected from the test suite described in Section \ref{sec:res-ensembles}.
We tested using a laptop with an Intel Core i7-2630QM CPU at 2.0 GHz and 8 GB memory.
This is a memory machine ratio of 512:1 (see Section \ref{sec:bg-scale})

The two performance tasks were:
(i) visualize time steps sequentially across a single run, and
(ii) visualize a single time step across multiple runs. %, and
Figure \ref{fig:vis-test-patterns} shows these test patterns in the context of the 2D ensemble grid.
These patterns were chosen to resemble natural exploration of an ensemble.

For each task, memory usage and disk read activity were recorded using the Python \texttt{memory\_profiler} module and the \texttt{iostat} Linux command line tool.
Before each test, disk cache was erased to remove any bias from previous testing.
The same tasks were performed with VisIt on the original data as a baseline for comparison.
A short script was written using VisIt's Python API to automate the process for each task.

\begin{figure}
    \centering
    \includegraphics[width=0.9\linewidth]{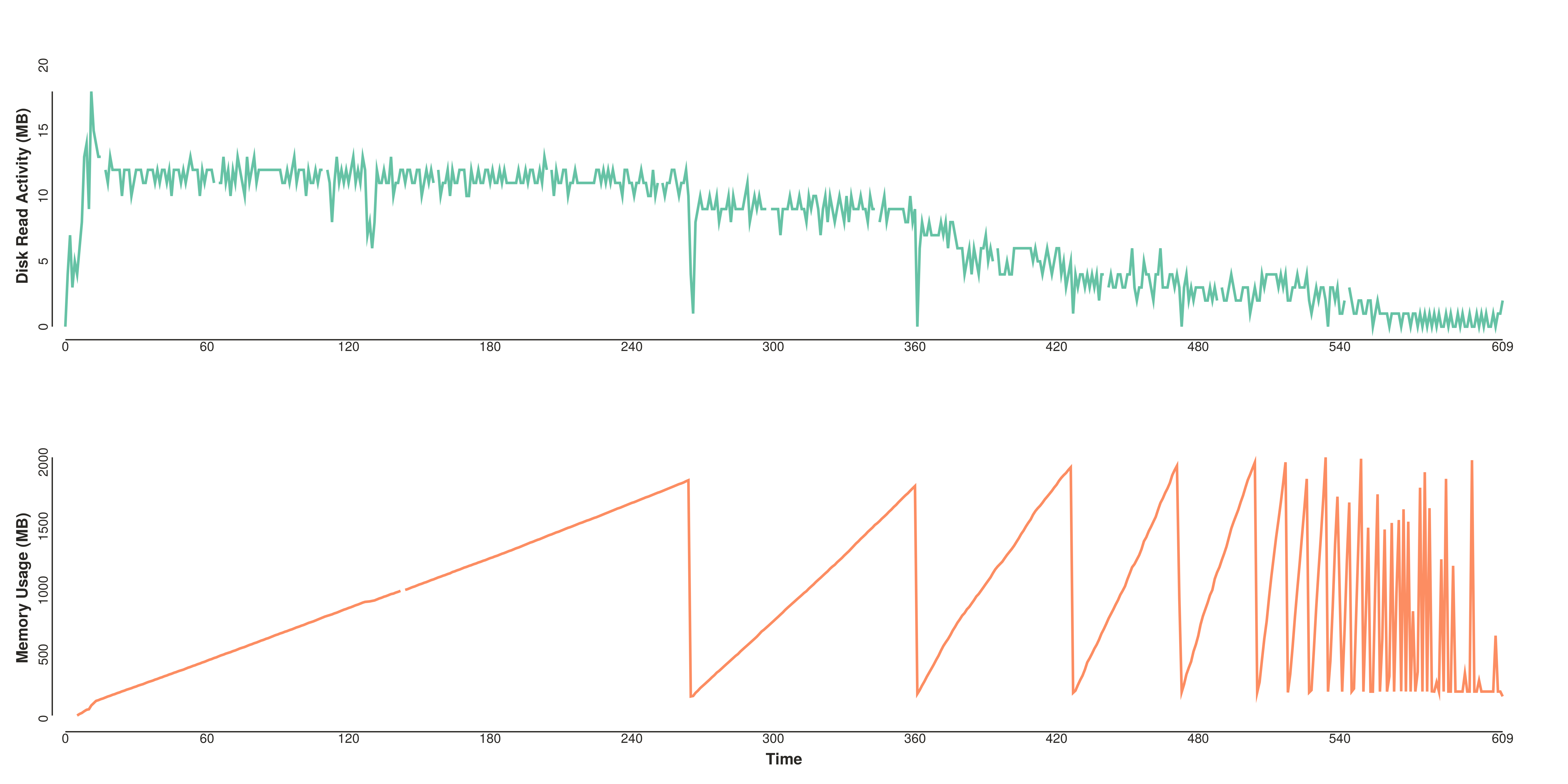}\\(a)\\
    \includegraphics[width=0.9\linewidth]{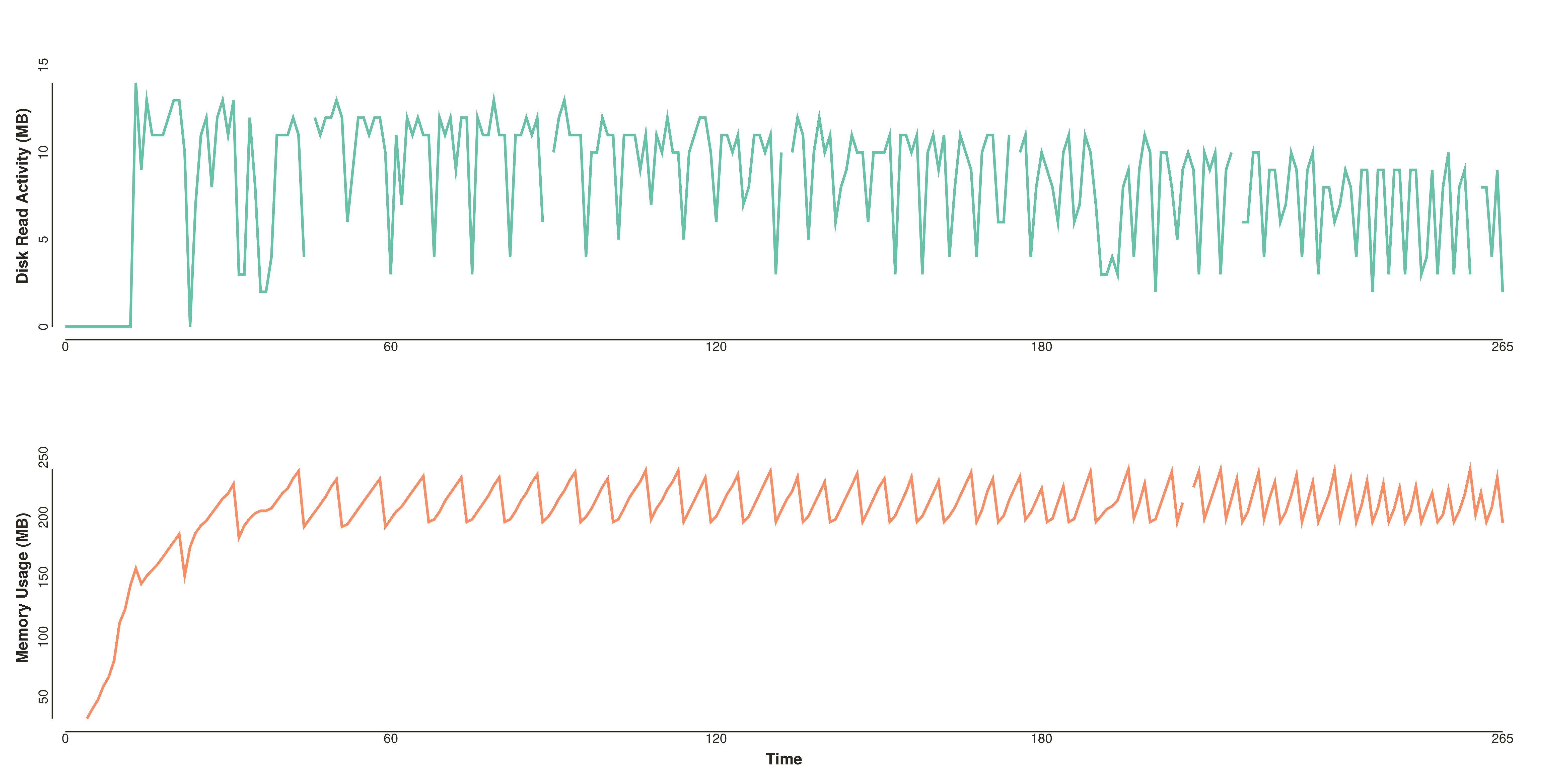}\\(b)\\
    \includegraphics[width=0.9\linewidth]{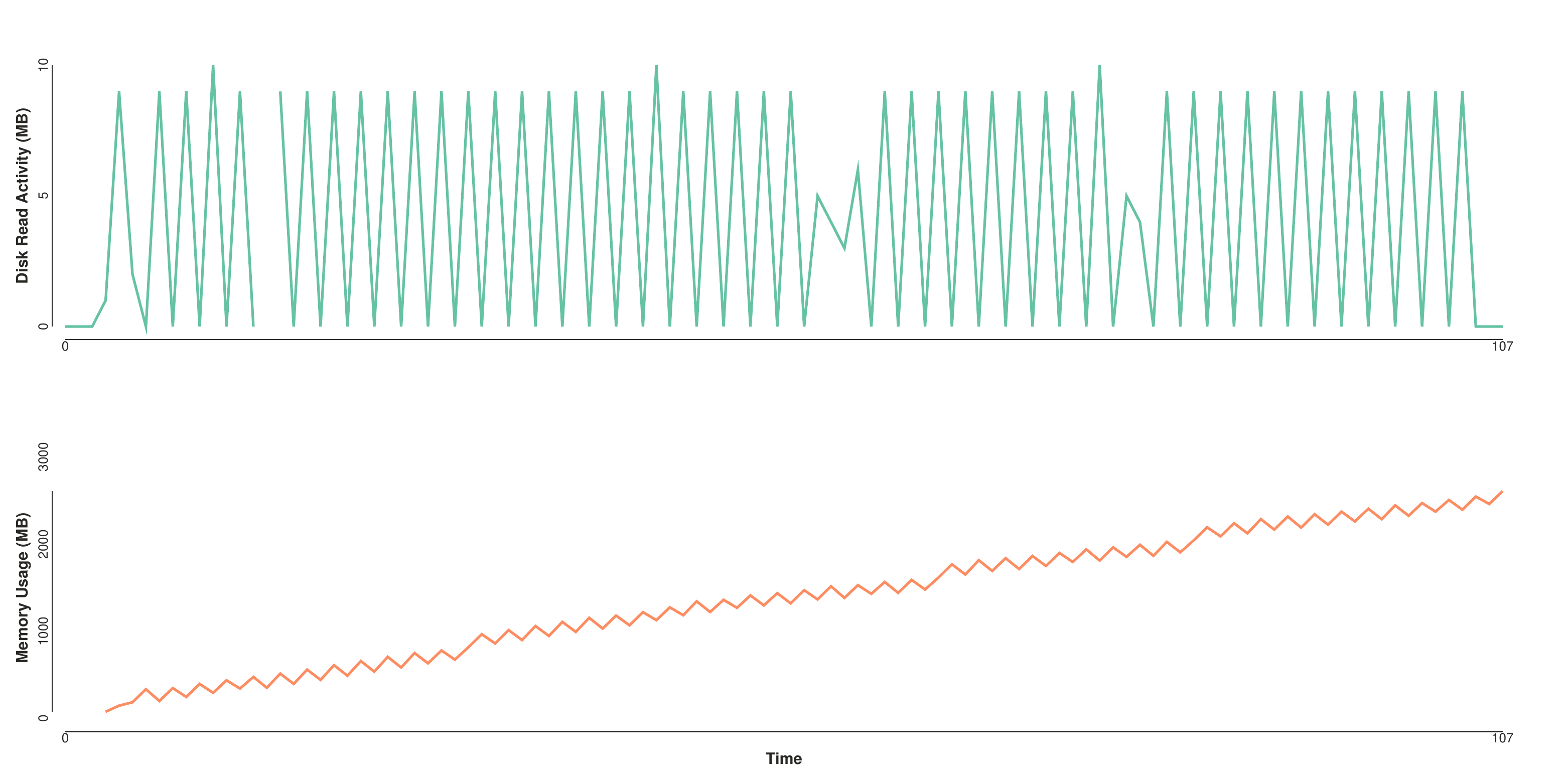}\\(c)\\
    \caption{Disk read activity and memory usage for the tasks described in Section \ref{sec:res-perf}. Plot (a) shows task (ii) for a codebook with $4\times 4\times 4$ block size. Initial reads were slow when seeking through the dense grid representation, but accelerated as the blocks remained in cache. Plot (b) shows task (i) for a codebook with $16\times 16\times 16$ block size. Access times are much quicker overall, but do not accelerate as dramatically as $4\times 4\times 4$. Plot (c) shows VisIt running task (i). Although completion time was faster, memory grew without bound.}
    \label{fig:perf-plots}
\end{figure}

Figure \ref{fig:perf-plots}a shows plots of disk activity and memory usage during task (ii) for a codebook with block size $4\times 4\times 4$, 1 decimal place, and 8 PCA components.
Memory usage is bounded at an upper limit, approximately 2 GB, determined by the block size, grid size, and degree of deduplication.
Note that as the disk cache is populated with blocks, visualization accelerates dramatically; the time between each spike in memory usage is successively shorter.

On the other hand, Figure \ref{fig:perf-plots}b shows task (i) for a codebook with block size $16\times 16\times 16$.
Though task (i) requires more volumes to be loaded (since there are 49 time steps as opposed to 40 runs), the test completed quicker than task (ii) due to less seek time in the codebook.
Additionally, there were fewer blocks per volume to load and reconstruct, allowing for quicker access from the codebook.

Figure \ref{fig:perf-plots}c shows VisIt running task (i).
Completion time was faster than the visualization component with a block size of $16\times 16\times$, but memory growth was unbounded.
With either block size in the visualization component, memory usage was bounded based upon the input parameters chosen.
The data segmentation also allows for disk cache to accelerate the visualization process.

In these tests, the primary measurement is load and reconstruction time.
In other words, how long it took to respond to an ensemble traversal request.
Once the disk cache had been populated, the smaller $4\times 4\times 4$ block size was capable of loading and reconstructing roughly one volume per second.
The $16\times 16\times 16$ block size codebook was able to load a volume in six seconds in the outset, and eventually reach one volume per three seconds as disk cache was populated.
This measurement is different from the traditional frames per second (fps) measure generally used for rendering performance.
With the visualization tool, the fps is not tied to the parameters used for the codebook.

\subsection{Profiling}\label{sec:res-profile}

We provide a heuristic to aid choosing parameters based on sampled regions of the data, since choosing optimal parameters for the NEA processing component may be challenging.
During testing, we chose a wide range of parameter values that allowed for exploration of the parameter space.
These heuristics are based on sampled regions of the dataset as processing the full dataset repeatedly is too expensive.
%Our performance tests showed a well-defined set of trade-offs based on the different input parameters.

\begin{figure}
    \centering
    \includegraphics[width=0.4\linewidth]{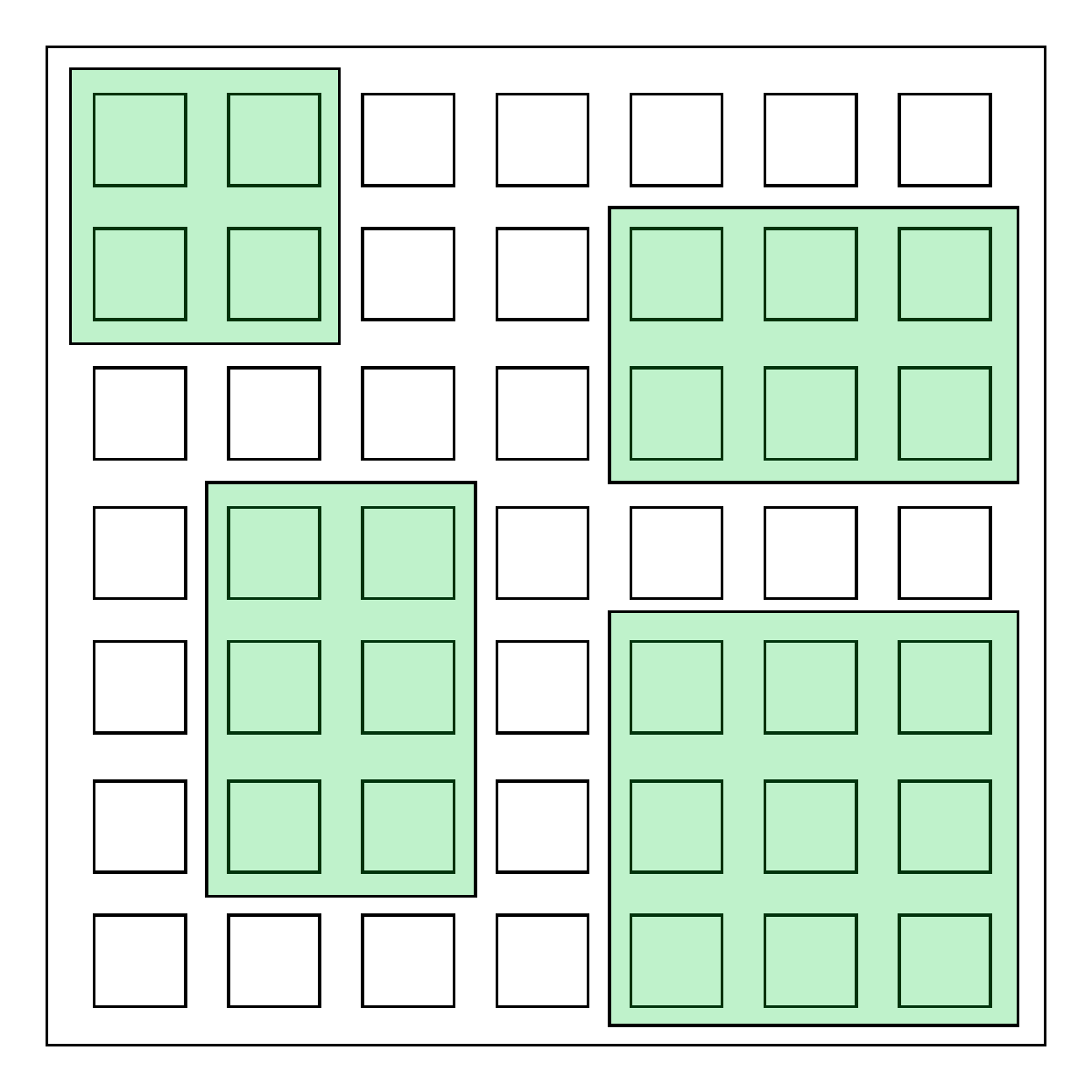}
    \caption{Example of sample regions randomly selected from the ensemble space when profiling. Regions range from $2\times2$ to $3\times3$ in size. Deduplication is done only on these regions to save time and provide an estimate of resultant codebook size with the given parameters.}
    \label{fig:ensemble-matrix-samples}
\end{figure}

We wrote a profiling script that provides an estimate for two values: the final codebook size, and the amount of memory required during visualization.
When profiling, the processing component takes several random non-overlapping samples from the ensemble space instead of processing the entire ensemble.
Sampled regions are randomly chosen from anywhere in the ensemble space, with a size ranging from $2\times 2$ to $3\times 3$ runs and time steps.
Figure \ref{fig:ensemble-matrix-samples} shows an example of sampled regions in the ensemble space, highlighted in green.
The number of samples taken is controllable; the default is to sample until at least 10\% of the ensemble space is covered.
These regions allow for deduplication across runs and time steps to be tested.

Only the volumes within the sampled regions are sent for a trial deduplication run.
After deduplicating, the final codebook size is estimated using the function
$$ S_{cb} = S_{od}\frac{B_{rem}}{B_{tot}}R + S_{g} $$
where $S_{od}$ is the size of the original data;
$S_{g}$ is the calculated size of the grid representation;
$B_{rem}$ and $B_{tot}$ are the number of blocks remaining after deduplication and the total possible number of blocks, respectively;
and $R$ is the estimated PCA reduction calculated as $$R=2\frac{E_{PCA}}{E_{tot}}$$ where $E_{PCA}$ is the number of PCA components kept, and $E_{tot}$ is the number of elements in a block.
The amount of memory required for visualization is estimated with the function
$$ M_{vis} = \frac{115200}{E_{tot}} + 200 $$
This function was determined by curve-fitting ($r^2=0.986$) the maximum memory recorded during the tests in Section \ref{sec:res-perf}.

The profiling script allows for large swaths of the parameter space to be tested quickly.
After the trial runs are complete, the best $n$ (configurable; 3 by default) results are shown.
From here, users may choose the set of parameters that provides the best codebook size and memory requirement to suit their machine.

\subsection{Other techniques}\label{sec:res-wavelet}

\begin{figure*}
    \centering
    \includegraphics[width=0.49\linewidth]{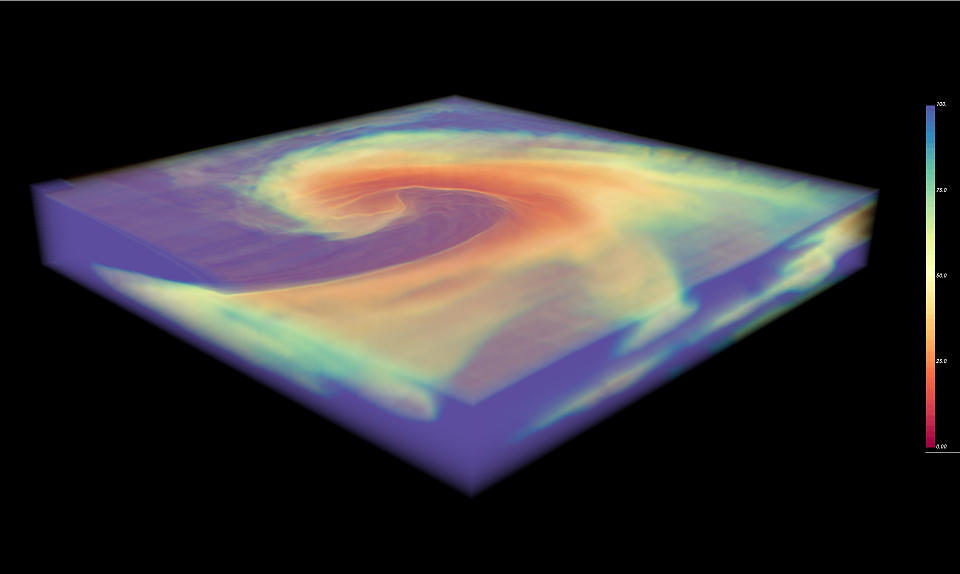}
    \includegraphics[width=0.49\linewidth]{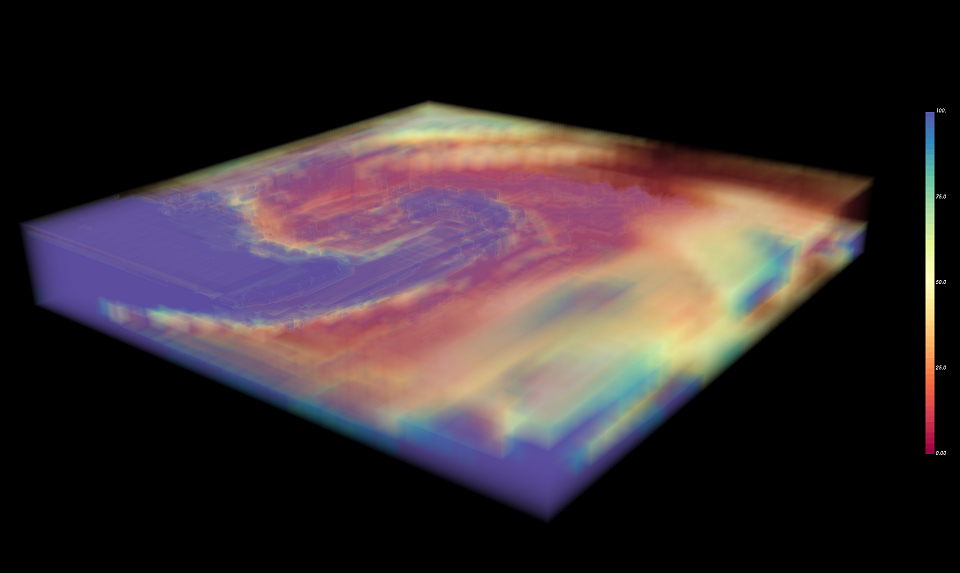}
    \caption{Two examples using wavelet compression instead of PCA for block-level data compression. The codebook on the left used a $16\times 16\times 16$ block size, 2 decimal place approximation, and 99\% wavelet quality. The codebook on the right used $8\times 8\times 8$ block size, -1 decimal places, and 50\% compression, a very aggressive parameter set.}
    \label{fig:wavelet-samples}
\end{figure*}

To demonstrate the use of another technique, we tested with wavelet compression instead of PCA.
Discrete wavelet transform (DWT) is commonly used to transform 2-dimensional image data, but is easily extended into 3 dimensions as it is a separable filter.
We use a standard implementation of wavelet compression consisting of wavelet transformation, thresholding and quantizing, and run-length encoding.

Our implementation uses the Haar wavelet for performance during compression and decompression.
This, however, limits block sizes to powers-of-2 dimensions.
The number of transformation levels for each block in dimension $d$ is then equal to $\log_2(d)$.
%We overwrite the block with its corresponding decomposed octree (Figure \ref{fig:wavelet-decomposition}), as is done with image transformation, to save memory.
Soft thresholding is applied using a scaled Universal Threshold, $\lambda^u_n = \sqrt{2\log(n)}\hat{\sigma}$, proposed by Donoho et al \cite{donoho1994ideal}, where $n$ is the number of data points or voxels.
The threshold is scaled by $\hat{\sigma} = 2\frac{100-q}{100}MAD$, where $q\in[0,100]$ is a user-supplied quality parameter and $MAD$ is the median absolute deviance of the wavelet coefficients, to allow for configurable compression strength.
After thresholding and quantization, wavelet coefficients are entropy encoded and run-length Huffman encoded into a byte stream using the encoding models proposed by Guthe and Stra\ss{}er \cite{Guthe2002}.

Figure \ref{fig:wavelet-samples} shows two example renders of wavelet compressed data.
The left codebook used a $16\times 16\times 16$ block size, 2 decimal place approximation, and 99\% wavelet quality, and compressed to 1.5 GB, an 11.6x reduction.
The right codebook used a much more aggressive parameter set: $8\times 8\times 8$ block size, -1 decimal places, and 50\% quality.
It compressed to 345 MB, a 51.65x reduction, albeit at great cost to data quality.
In general, wavelet compression provided much higher compression ratios even with very high quality parameters.

Most codebooks were under a gigabyte in size, showing over 17x reduction in size.
However, due to the quantization, there are always discontinuous artifacts present, especially between block borders.
This could be alleviated either with implementations of other wavelets during transformations, or with a different encoding method.
Wavelet compression and inverse transformation were a much more computationally demanding process.
Larger block sizes fared much better than smaller ones.
Due to the grid density with small block sizes, the inverse transform became unmanageable during visualization.

\subsection{Data Quality}\label{sec:res-quality}

To measure the variation of data quality as a function of codebook parameters, we measured the peak signal-to-noise ratio (PSNR) of various renders from the visualization component.
PSNR is a standard measurement technique for quality in the image processing and data compression fields.
Here, we use renders from a single viewpoint for four different codebooks that used strong compression parameters.

Table \ref{tab:psnr} shows the PSNR for renders from a single viewpoint for four different codebooks.
The chosen codebooks used stronger parameters to highlight the typical PSNR for heavily compressed codebooks.
Each render was of the last timestep from the first run of the ensemble. This is during a turbulent part of the storm, where most artifacts show up.
PCA performs fairly well as does a high quality wavelet transformation.

\begin{table}
    \centering
%    \small
    \begin{tabularx}{\linewidth}[X X r]{>{\hsize=1.5\hsize}X>{\hsize=0.5\hsize\raggedleft\arraybackslash}X>{\hsize=0.5\hsize}r}
        \toprule[1pt]
        Codebook configuration & Codebook size (GB) & PSNR (dB) \\
        \midrule
        PCA $16\times16\times16$ block size, 0 decimal places, 256 components & 2.8 & 28.27 \\
        PCA $4\times 4\times 4$ block size, 1 decimal place, 8 components & 4.9 & 27.63 \\
        Wavelet $16\times 16\times 16$ block size, 2 decimal places, 99\% quality & 1.5 & 29.74 \\
        Wavelet $8\times 8\times 8$ block size, -1 decimal places, 50\% quality & 0.34 & 20.86 \\
        \bottomrule[1pt]
    \end{tabularx}
    \caption{PSNR values for renders of various codebooks compared to the original data. The same viewpoint was used for each render. The codebooks presented here used stronger parameters choices. In each case, the last timestep from the first run was used.}
    \label{tab:psnr}
\end{table}

\section{Conclusion and Future Work}\label{sec:conclusion}

In this paper, we presented NEA, a navigable ensemble analysis and visualization tool.
Ensemble analysis has been a long-standing challenge for the visualization community, as ensemble data push current algorithms with their size and complexity.
We have shown that using a new \textit{codebook} structure, NEA can manage the working set of an ensemble visualization session with a fine degree of control to enable interactive visual exploration of ensembles using commodity hardware.
Our approach combines existing methods of data segmentation and duplicate removal with data reduction.

For future work, we would like to compare the use of locality-sensitive hash algorithms to the current quantize-and-hash steps, as well as several different data compression methods for block-level reduction.
Adaptive compression techniques may be used as well.
Our implementations for PCA and wavelet both use a fixed and known transformation depending on the parameters chosen.
Instead, PCA components may vary on a per-block or region basis.
Additionally, the pipeline could be used in coordination with in-situ visualization.
This would allow for processing data at simulation time to save only a reduced form of the data that can be explored post-hoc.

\section*{Acknowledgments}
The authors are supported in part by NSF Award CNS-1629890, USDI-NPS P14AC01485, Intel Parallel Computing Center (IPCC) at the Joint Institute of Computational Science of University of Tennessee, and the Engineering Research Center Program of the National Science Foundation and the Department of Energy under NSF Award Number EEC-1041877.
Mississippi State University provided the WRF simulation data.
The authors would like to thank the anonymous reviewers for their valuable comments and suggestions.

\bibliographystyle{eg-alpha-doi}
\bibliography{nea}

\end{document}